\documentclass[amsmath, floatfix, superscriptaddress]{revtex4-2}
\usepackage{physics}
\usepackage{siunitx}
\usepackage{graphicx}
\usepackage{microtype}
\usepackage[T1]{fontenc}
\usepackage{soul,color}

\begin{document}

\title{Resonance-free deep ultraviolet to near infrared supercontinuum generation in a hollow-core antiresonant fibre}

\author{Mohammed Sabbah}
\affiliation{School of Engineering and Physical Sciences, Heriot-Watt University, Edinburgh, EH14 4AS, United Kingdom}
\author{Robbie Mears}
\author{Kerrianne Harrington}
\author{William J. Wadsworth}
\author{James M. Stone}
\author{Tim A. Birks}
\affiliation{Centre for Photonics and Photonic Materials, Department of Physics, University of Bath, Claverton Down, Bath, BA2 7AY, United Kingdom}
\author{John C. Travers}
\email{j.travers@hw.ac.uk}
\affiliation{School of Engineering and Physical Sciences, Heriot-Watt University, Edinburgh, EH14 4AS, United Kingdom}

\begin{abstract}
Supercontinuum generation in the ultraviolet spectral region is challenging in solid-core optical fibres due to solarization and photodarkening. Antiresonant hollow-core fibres have overcome this limitation and have been shown to guide ultraviolet light at sufficient intensity for ultraviolet spectral broadening through nonlinear optics in the filling gas. However, their ultraviolet guidance is usually limited by discontinuities caused by the presence of high-loss resonance bands. In this paper, we report on resonance-free supercontinuum generation spanning from the deep ultraviolet to the near infrared achieved through modulation instability in an argon-filled antiresonant hollow-core fibre. The fibre was directly fabricated using the stack-and-draw method with a wall thickness of approximately 90~nm, enabling continuous spectral coverage from the deep ultraviolet to the near infrared. We also report numerical simulations to investigate the supercontinuum bandwidth and the factors limiting it, finding that the overall dispersion landscape, and associated group-velocity matching of cross-phase modulation interactions, is the dominant constraint on spectral extension.
\end{abstract}

\maketitle

\section{Introduction}
\noindent Supercontinuum generation is a well-established technique that converts laser emission into an extremely broad bandwidth with high brightness, high spatial coherence, and sometimes high temporal coherence~\cite{Dudley2006, Genty2007, alfano_supercontinuum_2016, Sylvestre:21}. In silica optical fibres, supercontinuum sources can span from the infrared to the near ultraviolet (UV)~\cite{Travers_2010}; however, further extension into the ultraviolet spectral region is constrained by optical damage due to photodarkening and solarization effects in the glass core~\cite{Skuja2001}. Alternative soft-glass fibres, such as ZBLAN, have been explored for ultraviolet supercontinuum generation~\cite{jiang2015}, however, these fibres are very challenging to fabricate and not widely available. Hollow-core antiresonant fibres have emerged as a promising alternative for supercontinuum generation and frequency conversion that can extend into the deep and even vacuum ultraviolet region~\cite{Belli:15, Ermolov2015, Yu:18, hosseini2018uv, smith2020low, Suresh2021, Mears:24}. Resonant-dispersive wave (RDW) generation has been the predominant technique for supercontinuum generation in the UV spectral region using hollow-core antiresonant fibres~\cite{joly2011bright, Mak2013, Belli:15, Ermolov2015, Kottig2017, hosseini2018uv, smith2020low}. This approach typically requires input pulses with durations of a few tens of femtoseconds or less to seed coherent soliton self-compression. A less explored technique in hollow-core fibres is modulational instability (MI)~\cite{Travers:11}. In MI-driven supercontinuum generation, high-power pump pulses with relatively long duration amplify noise-seeded parametric spectral sidebands, corresponding to pulse breakup and the formation of a shower of few-cycle solitons in the time domain~\cite{Hasegawa1986, Dudley2006, Tani2013}. The resulting solitons can also drive RDW emission, enhancing energy transfer to the normal dispersion region, and supercontinua extending into the ultraviolet have been demonstrated with this technique using antiresonant hollow-core fibres~\cite{Tani2013, Sabbah2023}.  As an alternative approach, significant attention has recently been focused on Raman-enhanced supercontinuum generation with molecular gases~\cite{Belli:15,mousavi_nonlinear_2018,hosseini2018uv,gao_raman,Sabbah2023,gladyshev2024mid,plosz2024}, often providing very flat and broad supercontinua. However, apart from the soliton-driven works of Belli~\textit{et al.}~\cite{Belli:15} and Hosseini~\textit{et al.}~\cite{hosseini2018uv}, which are often not very flat, molecular gas supercontinua have not been extended to the deep ultraviolet so far.

A critical consideration when using antiresonant hollow-core fibres is the locations of the high-loss bands at the resonances intrinsic to their guidance mechanism. These resonances interrupt the fibre transmission, and can considerably disrupt or constrain the continuity, extent and flatness of the supercontinuum, affecting almost all supercontinuum results in antiresonant fibres reported so far. The locations of the resonances are determined by the thickness of the core wall $t$, and are given by $\lambda_p \approx 2t(n_\mathrm{glass}^2-1)^{1/2}/p$, where $n_\mathrm{glass}$ is the glass refractive index, and $p$ is the order of the resonance~\cite{Yu2016}. The widest low-loss transmission bandwidth---required for supercontinuum generation---is achieved in the fundamental band ($p < 1$), requiring the thinnest walls. The wall thickness should be chosen to be $t \sim \lambda_0/4$, where $\lambda_0$ is the wavelength at the middle of the transmission band. The core wall thickness can be optimized for a specific application, from a few microns thick for mid-infrared guidance to hundreds of nanometers thick for the visible and ultraviolet. To achieve resonance-free transmission from 200~nm in the deep ultraviolet to the near infrared, the thickness of the core wall must be less than \(\sim\)90~nm. Achieving such thin core walls is challenging, and has previously required post-processing techniques such as etching and tapering~\cite{Tani2018, hosseini2018uv, Hoang2021, Winter2019}. As far as we are aware, the thinnest core-wall fabricated directly on the drawing tower was achieved by Ding~\textit{et al.}~\cite{Ding2020}, who obtained $\sim \SI{114}{\nm}$. 

Hosseini~\textit{et al.}~\cite{hosseini2018uv} and Suresh~\textit{et al.}~\cite{Suresh2021} have both demonstrated supercontinuum generation extending down to the deep ultraviolet (at the \(\sim 15\)~dB level or lower) by post-processing an antiresonant hollow-core fibre. In the case of Hosseini~\textit{et al.} the post-processing was used to shift the fibre resonances away from the supercontinuum region. In the case of Suresh~\textit{et al.} the down-tapering transition was utilised to enhance the nonlinear dynamics, and the mechanism suggested for the spectral extension was that the tapering induced a shift in the zero-dispersion wavelength (ZDW) toward shorter wavelengths, thereby enabling phase-matching for MI-driven RDW generation to extend further into the deep-UV (\(\sim\)213 nm).

In this paper, we report on the generation of a smooth and continuous supercontinuum extending from the deep ultraviolet to the near infrared by utilising an ultrathin-walled single-ring antiresonant hollow-core fibre designed such that it is resonance-free across the whole spectral region. Removing the resonances enables enhanced supercontinuum efficiency and flatness. In contrast to previous work, our fibre was directly fabricated to its final geometry by the stack-and-draw technique, without the need for post-processing of the fibre by tapering or etching. We achieved a core-wall thickness of approximately \(\sim\)\SI{90}{\nm}, which is, to the best of our knowledge, the thinnest antiresonant fibre core-wall fabricated directly on the drawing tower. This approach simplifies fabrication and enables the production of long uniform fibre lengths, as are often necessary for MI-based supercontinuum (in the present work we used a 75~cm length), and are useful for pulse delivery \cite{Shephard:04, Lekosiotis:23}. The resonance-free supercontinuum we generated covers the spectral range from \SI{260}{\nm} to \SI{750}{\nm} while maintaining excellent spectral flatness and efficiency across the entire bandwidth. We use numerical simulations to explore the factors limiting the bandwidth of the supercontinuum generation in our experiments. By employing the capillary dispersion model we match our experimental results and investigate the impact of higher-order modes and soliton-dispersive wave interaction (through cross-phase modulation (XPM)) on the spectral broadening.

\section{Experiment}
\subsection{Fibre}
\begin{figure}[b!]
\centering
\includegraphics{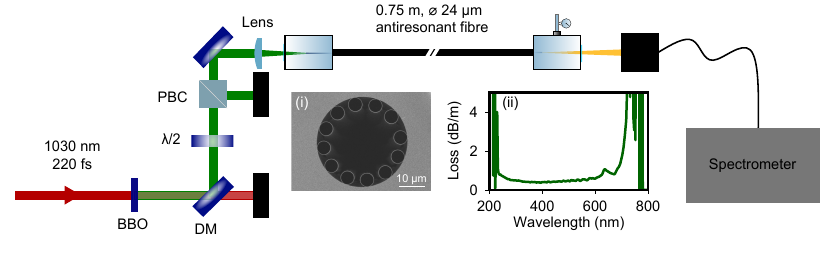}
\caption{Experimental setup. DM: dichroic mirror; $\lambda/2$: half-waveplate; PBC: polarizing beam cube. The insets show: (i) a scanning electron microscope image of the fibre cross-section; (ii) the measured loss profile.}
\label{fig}
\end{figure}
\noindent Our ultrathin core-wall fibre was fabricated using the stack-and-draw method, using a single intermediate cane stage. A cane of $\sim\SI{2.2}{\mm}$ diameter was jacketed to form a 3.4~mm outer diameter preform and drawn to a fibre outer diameter of \SI{77}{\um} under a drawing tension of 190~g. A small draw-down ratio and high drawing stress allowed a long uniform length of 185~m to be drawn, with high structural uniformity despite the extremely thin wall. The resulting core diameter was \SI{24}{\micro\meter}. An optical cut-back measurement using a broadband incoherent light source exciting all modes (note that this technique provides a strict upper bound on the loss of the fundamental mode) indicated a minimum loss of 0.5~dB/m at 450~nm. The full measured loss curve is shown in Fig.~\ref{fig}(ii). The estimated wall thickness of \(\sim\)90~nm was calculated based on the observed 200~nm wavelength of the first resonance ($p = 1$) and volume conservation during fabrication.

\subsection{Setup}

\noindent Figure~\ref{fig}(a) shows the experimental setup. As the guidance band of our fibre was optimised for the deep-ultraviolet to visible range we chose to pump the fibre at 515~nm through second-harmonic generation (SHG) of our Yb pump laser. Pumping via SHG offers the additional advantage of more efficient energy transfer to the deep ultraviolet region~\cite{Sabbah2023, Sabbah2024}. We frequency doubled our Yb:KGW pump laser (\SI{1030}{\nm} wavelength, \SI{220}{\fs} full-width half-maximum pulse duration) in a \SI{3}{\mm} thick BBO crystal to \SI{515}{\nm}. The residual \SI{1030}{\nm} was removed using a pair of dichroic mirrors. The beam was then attenuated using a half-wave plate and a polarisation beam cube and coupled into the \SI{24}{\micro\meter} diameter core of the fibre using a plano-convex lens. The \SI{75}{\cm} fibre was placed between gas cells and filled with argon at different pressures with optical access provided by a pair of \SI{3}{\mm} thick uncoated fused silica windows. The output spectrum was collected using an integrating sphere connected to a fibre-coupled CCD spectrometer. The spectrometer covers the spectral range 200~nm to 1160~nm (Avantes ULS2048XL). The whole system is calibrated on an absolute scale with NIST traceable lamps. High dynamic range spectra were separately collected using an integrating sphere coupled to a scanning double monochromator. This system was also calibrated as a whole for relative spectral response.

\subsection{Results}
\begin{figure}[b!]
\centering
\includegraphics{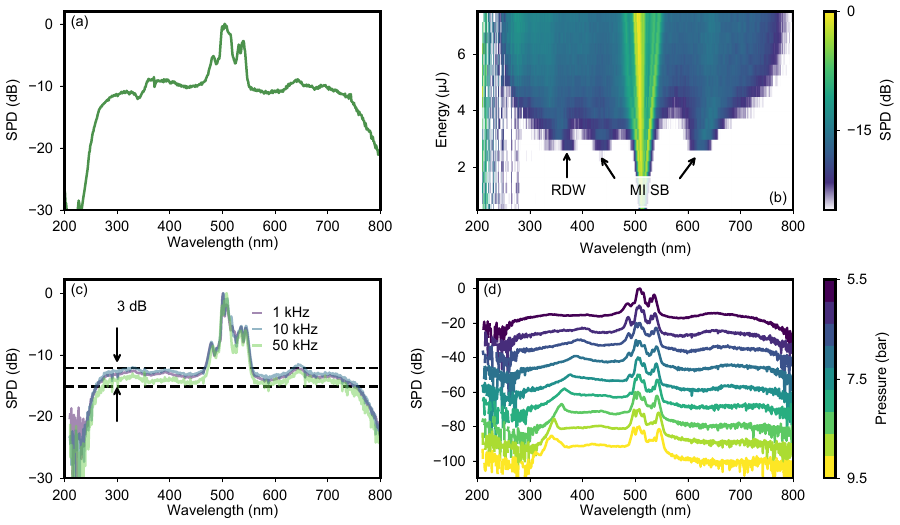}
\caption{(a) Experimental supercontinuum spectrum generated with a \SI{5.5}{\bar} argon filling pressure and \SI{6}{\micro\joule} pump energy. (b) Evolution of the output spectrum with a \SI{5.5}{\bar} argon filling pressure as the input energy is increased. The arrows indicate the position of RDW and MI sidebands (SB). (c) Experimental output spectrum at several repetition rates, with a \SI{5.5}{\bar} argon filling pressure and \SI{6}{\micro\joule} pump energy, normalized to the 1~kHz case. The dashed horizontal lines indicate the flat part of the supercontinuum with 3~dB level flatness. (d) Experimental output spectrum as the filling pressure increases from \SI{5.5}{\bar} to \SI{9.5}{\bar} at \SI{5}{\micro\joule} pump energy.}
\label{fig2}
\end{figure}

\noindent Figure~\ref{fig2}(a) illustrates an example of the output supercontinuum generated when the fibre is filled to \SI{5.5}{\bar} of argon, with an input energy of approximately \SI{6}{\micro\joule}.  In this case, the ZDW is at 469~nm, the pump soliton order is approximately 110, and the pump wavelength (\SI{515}{\nm}) lies within the anomalous dispersion region. The resulting dynamics are dominated by MI. The generated supercontinuum spans the first transmission window of the fibre, covering a wavelength range from \SI{260}{\nm} to \SI{750}{\nm} (at the 3~dB level), and exhibits excellent spectral flatness across this continuum, except for the unconverted pump region, which we show below to be residual higher-order mode content that can be eliminated in future work. Note that the difference between Figure~\ref{fig2}(a) and the rest of the figures using the same parameters is due to the saturation of the spectrometer around 515~nm. This was deliberate to show the full bandwidth extension of the spectrum at a higher dynamic range.

Figure~\ref{fig2}(b) shows the experimental output spectral evolution as the input energy increases when the fibre is filled with \SI{5.5}{\bar} of argon. Initially, the spectrum undergoes self-phase modulation up to approximately \SI{2.5}{\micro\joule}. Then two MI sidebands appear before the flat supercontinuum develops. The calculated MI sideband wavelengths are located at 422~nm and 660~nm, which closely match the measured values. A distinct spectral peak appears in the ultraviolet region around \SI{370}{\nm}, generated by RDW emission from solitons, as discussed in~\cite{Suresh2021, Sabbah2023}. However, this peak is less pronounced compared to other studies, likely due to its location within the MI gain bandwidth. The calculated wavelength for the RDW is 390~nm, which is in reasonable agreement with the measured spectrum.

Scaling the repetition rate from 1~kHz up to 50~kHz results in minimal change to the output spectrum, as shown in Figure~\ref{fig2}(c). At 50~kHz, the optical power across the flat part of the spectrum is around 70~\SI{}{\micro\watt/ \nm}. The current experiments were limited by the repetition rate of our pump laser system.

Figure~\ref{fig2}(d) shows how the output spectrum changes with pressure using a fixed input energy. The supercontinuum bandwidth reduces as the pressure increases, while the spectral power density is larger in the near UV with increasing pressure. At higher pressures, the supercontinuum bandwidth is limited because the low-frequency MI sideband falls within the fibre's high-loss region, above 800~nm. For example, at 9.5~bar argon and using \SI{5}{\micro\joule} pulse energy, the calculated central wavelength for the first MI sideband is at \SI{1.2}{\micro\meter}. This significantly limits the spectral power density (SPD) above the pump wavelength. We attributed the prominent peak around 360~nm for 9.5~bar pressure (and similar peaks at lower pressures) to the propagation of higher-order modes, and confirmed this by observing the spectrally filtered far-field profile (not shown).

\begin{figure}[b!]
\centering
\includegraphics{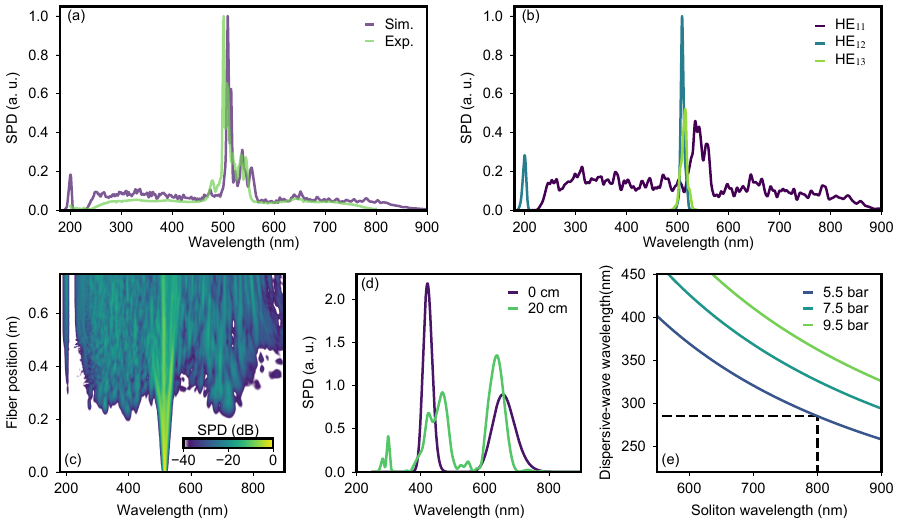}
\caption{(a) Comparison between experiment and simulation for 5.5~bar argon and \SI{6}{\micro\joule} input pulse energy. The simulated spectrum includes the first four HE$_{1m}$ modes. We coupled 85\% of energy to the fundamental mode and the remainder was distributed equally between the first three higher order modes. The calculated spectra are processed to reproduce a 5~nm spectral resolution and averaged over 50 simulations, each with a different noise seed~\cite{Drummond2001}. (b) The simulation output spectra from (a) for the first three HE$_{1m}$ modes. (c) Simulated spectral evolution along the fibre for all modes.  (d) Simulation for a single soliton ($\mathrm{energy} = \SI{78}{\nano\joule}$, $\tau_\mathrm{fwhm} = \SI{7.8}{\fs}$, $\lambda = \SI{660}{\nm}$) and a single dispersive wave ($\mathrm{energy} = \SI{78}{\nano\joule}$, $\tau_\mathrm{fwhm} = \SI{7.8}{\fs}$, $\lambda = \SI{422}{\nm}$) showing the effect of XPM using 5.5~bar argon. (e) Group-velocity
matching curves for three different pressures.}
\label{fig3}
\end{figure}

\section{Simulations and Discussion}

\noindent We performed numerical simulations using \textit{Luna.jl}~\cite{brahms_lunajl_2021} to identify the limiting factors for the supercontinuum bandwidth. The capillary model~\cite{Marcatili1964} was used to calculate the dispersion, as it gives results that closely match our experimental results, which is expected when considering only the first transmission window of an antiresonant fibre. Fibre attenuation was neglected. The simulations included the first four higher-order modes of type HE$_{1m}$, where \(m=1,2,3,4\). We select these modes as they are most strongly involved in nonlinear dynamics when the fibre is excited in the fundamental mode.

Figure~\ref{fig3}(a) provides a comparison between the experimental output spectrum and the simulated spectrum for the same experimental parameters as in Figure~\ref{fig2}(a), averaged over 50 simulations with different noise seeds~\cite{Drummond2001}. The simulation contains contributions from all modes. We attribute the observed discrepancies between the two spectra to fibre loss (neglected in the simulations). However, the fairly good agreement in terms of spectral extension, at both long and short wavelengths, between the loss-free modelling and experiment suggests that the fibre transmission bandwidth has not severely restricted the supercontinuum extent.

Figure~\ref{fig3}(b) presents the numerically calculated output spectrum for the first three HE$_{1m}$ modes individually. The higher-order modes remain largely unaffected and make minimal contributions to the supercontinuum generation. They remain around the pump wavelength, causing the spike around 515~nm observed in both the experiment and simulations in Figure~\ref{fig3}(a). The fundamental mode forms the supercontinuum. In the simulations, it also phase-matches and emits a resonant dispersive wave in a higher-order mode, as evidenced by the peak at 200~nm. This feature, however, is absent in the experimental results, likely due to a significant fibre loss in the higher-order modes not accounted for in the numerical model.

Figure~\ref{fig3}(c) shows the numerically calculated spectral evolution along the fibre. Initially, the supercontinuum forms within the MI gain bandwidth and, following the usual MI supercontinuum dynamics~\cite{Dudley2006}, consists of solitons and dispersive waves. The short wavelength extension is initially limited by the phase-matched RDW spectrum. Subsequently, a slight enhancement of the supercontinuum bandwidth is observed between 0.3~m and 0.5~m, which we attribute to XPM between colliding solitons and dispersive waves~\cite{Nishizawa:02,genty_route_2005, beaud_ultrashort_1987, skryabin_theory_2005,Travers:09,Sabbah2023}. This interaction is enhanced while the group velocity mismatch between the solitons and dispersive waves is small. It should be noted that we do not expect full RDW trapping by solitons in the present work, because we have neither Raman scattering nor axial waveguide variation, either of which is required to decelerate the solitons and enable trapping~\cite{beaud_ultrashort_1987,skryabin_theory_2005,Travers:09}.

We numerically investigated the effect of XPM on the dispersive wave spectrum. The numerical simulation involved a single soliton and a single dispersive wave pulse that were initially temporally overlapped. The soliton's energy and duration were estimated from the MI period, $E_{\text{sol}} = P_0 T_{\text{MI}}$, as discussed in Ref.~\cite{Tani2013}. The initial spectral locations of the soliton and dispersive-wave were estimated from the full numerical simulations. Figure~\ref{fig3}(d) presents the input and output simulated spectrum, illustrating the significant impact of XPM on the dispersive wave. The dispersive wave spectrum (initially centred at 422~nm) undergoes substantial modification, with the generation of new spectral components at shorter wavelengths. In contrast, the soliton's spectrum (initially centred at 660~nm) remains relatively unchanged. In the experiment, several tens of solitons and dispersive waves undergo a similar process through collisions which further enhances the supercontinuum extension to the deep ultraviolet. The interactions continue as long as the group-velocity mismatch between the solitons and dispersive waves remains small. To examine this, we calculated the group-velocity matching between the solitons and dispersive-waves. Figure~\ref{fig3}(e) shows group-velocity-matched pairs of wavelengths for three different pressures. We observe that at lower pressures the group velocity remains matched to shorter wavelengths. For instance, at 9.5~bar, 800~nm has roughly the same group velocity as 360~nm, while at 5.5~bar, 800~nm matches the group velocity of 285~nm. This observation aligns reasonably well with the experimental results, where fibre loss restricts the longest wavelength to around 800~nm, and our short-wavelength edge extends down to $\sim\SI{260}{\nm}$ in the case of 5.5~bar filling pressure.

\section{Conclusion}
We have demonstrated a resonance-free supercontinuum spanning from the deep ultraviolet to the near infrared spectral region in an antiresonant hollow-core fibre with ultrathin core wall thickness fabricated directly using the stack and draw method. We also numerically modeled the supercontinuum dynamics and found that they could be mostly reproduced without including the fibre loss. This suggests that it is the nonlinear dynamics and dispersion landscape---through the group velocity matching of XPM interactions---that are the main limiting factor for the spectral bandwidth, with the fibre loss playing a secondary role.

The development of long lengths of ultrathin core wall antiresonant fibres, drawn directly to fibre, opens the door to enhanced gas-based nonlinear optics in the ultraviolet spectral region. The supercontinuum demonstrated here should find great utility in sensing and metrology in the ultraviolet-visible region.

\begin{flushleft}
\noindent\textbf{Acknowledgments.} The authors thank Christian Brahms and Jonathan Knight for their useful discussions.
\end{flushleft}

\noindent\textbf{Funding.} This work was funded by the United Kingdom's Engineering and Physical Sciences Research Council: Grant agreement EP/T020903/1. JCT is supported by a Chair in Emerging Technology from the Royal Academy of Engineering and by the Institution of Engineering and Technology (IET) through the IET A F Harvey Engineering Research Prize.\linebreak

\bibliography{bibliography}

\end{document}